\documentclass[12pt]{article}
\usepackage{geometry}
\usepackage{amstext}
\usepackage{amssymb}
\usepackage{babel}
\usepackage{color}
\usepackage[colorlinks,allcolors=red]{hyperref}

\begin{document}
\title{Comment on a proposal for non-relativistic transverse electromagnetic Hamiltonian. }

\author{Ladislaus Alexander B\'anyai}

\maketitle
{\sl Institut f\"ur Theoretische Physik, Goethe-Universit\"at, Frankfurt am Main  
}
\\

{\bf banyai@itp.uni-frankfurt.de} {\color{blue}corresponding author} 

ORCID ({\color{blue}https://www.orcid.org/0000-0002-0698-4202)}

\abstract{We comment on the recent proposal (arXiv:2608.21211v1 (2026))of a new transverse Hamiltonian for electromagnetic
interacting charged particles that ignores both the classical and quantum-mechanical
versions of the already known one.}

\vspace{1cm}

\section{Introduction}
In a recent paper Marini, Wu and Reho {\cite{Marini}  try to construct a new Hamiltonian of electromagnetic interacting charged particles including transverse electromagnetic interactions. All previous approaches they quote are the  classical theory of Darwin \cite{Darwin} and the more recent quantum mechanical formulation of Bányai {\cite{Banyai} based on the old paper of Holstein, Norton and Pincus {\cite{Holstein}. These are thereafter fully discarded by a single sentence under the accusation of lacking gauge invariance. The accusation is at least strange, since these theories are built explicitly within the Coulomb gauge, therefore no gauge freedom remains.

The problem of the non-relativistic transverse electromagnetic interactions between charged  particles was however discussed in many papers as well as textbooks {\cite{Landau},\cite{Jackson}, \cite{Banyai1}. It implies  intricate mathematical aspects (see   Dirac’s canonical formalism \cite{Dirac1},\cite{Dirac2}) since due to the velocity
dependent terms, there is a relationship between the canonical momenta. A simple way to avoid this is to eliminate the spurious degree of freedom by the choice of the Coulomb gauge. Next one proceeds with  the common assumption of a development in powers of $ \frac{1}{c} $. An analysis of the Feynman diagrams according to  Holstein, Norton and Pincus {\cite{Holstein} leads in the subspace of states without photons to a current-current Coulomb-like interaction of the charged particles of order $\frac{1}{c^2}$ within the  non-relativistic QED
    
\[
-\frac{1}{2} \int d\vec{x}\int d\vec{x'}
\frac{N\left[\vec{i}_\bot(\vec{x})\vec{i}_\bot(\vec{x}')\right]} 
{c^2|\vec{x}-\vec{x}'| }\enspace 
\]
with $\vec{i}_\bot(\vec{x})$ being the transverse current density. This is an exact result without any supplementary assumptions beyond the already mentioned ones.   An analogous classical expression is obtained by Refs.{\cite{Darwin}, \cite{Landau}, \cite{Jackson}}, while the normal product there is replaced by omitting the divergent self-interaction of point-like charges. 

 The above current density-current density interaction  has the same form as the charge density-charge density Coulomb interaction.  It is the microscopical expression  of the important Biot-Savart law of  interaction between currents in finite macroscopic samples, where light propagation effects may be ignored. 

The only remaining aspect related to gauge invariance is the introduction of an external (classical)  vector potential. It follows in the trivial way by the minimal e.m, coupling as it was mentioned explicitly in earlier papers \cite{Banyai}, 
\cite{ Banyai-Bundaru}.

Any way, gauge invariance may be formulated only with classical fields. In the relativistic QED it is implemented through its functional integral formulation.

\section{Conclusions}
The paper of Marini,Wu and Reho {\cite{Marini} has to consider all the subtleties of the problem and compare their resulting transverse Hamiltonian with the above one. Differences may occur only due to other ingredients (assumptions) and have to be carefully commented.



\end{document}